\documentclass[%
 aps,
 prl,%
 amsmath,amssymb,
 superscriptaddress,
reprint,%
]{revtex4-1}

\usepackage{graphicx}
\usepackage{dcolumn}
\usepackage{bm}
\usepackage[utf8]{inputenc}
\usepackage{physics}

\begin{document}


\title[Sample title]{Probe of Multi-electron Dynamics in Xenon\\by Caustics in High Order Harmonic Generation
}
\author{D. Faccial\`a}
\affiliation{Dipartimento di Fisica, Politecnico di Milano, Milan, Italy}
\affiliation{Istituto di Fotonica e Nanotecnologie - CNR, Milan, Italy}
\author{S. Pabst}
\affiliation{Center for Free-Electron Laser Science DESY, Hamburg, Germany}
\affiliation{ITAMP Harvard-Smithsonian Center for Astrophysics, Cambridge (MA), USA}
\author{B. D. Bruner}
\affiliation{Department of Physics of Complex Systems, Weizmann Institute of Science, Israel}
\author{A. G. Ciriolo}
\affiliation{Dipartimento di Fisica, Politecnico di Milano, Milan, Italy}
\affiliation{Istituto di Fotonica e Nanotecnologie - CNR, Milan, Italy}
\author{S. \surname{De Silvestri}}
\affiliation{Dipartimento di Fisica, Politecnico di Milano, Milan, Italy}
\affiliation{Istituto di Fotonica e Nanotecnologie - CNR, Milan, Italy}
\author{M. Devetta}
\affiliation{Istituto di Fotonica e Nanotecnologie - CNR, Milan, Italy}
\author{M. Negro}
\affiliation{Istituto di Fotonica e Nanotecnologie - CNR, Milan, Italy}
\author{H. Soifer}
\affiliation{Department of Physics of Complex Systems, Weizmann Institute of Science, Israel}
\author{S. Stagira}
\affiliation{Dipartimento di Fisica, Politecnico di Milano, Milan, Italy}
\affiliation{Istituto di Fotonica e Nanotecnologie - CNR, Milan, Italy}
\author{N. Dudovich}
\affiliation{Department of Physics of Complex Systems, Weizmann Institute of Science, Israel}
\author{C. Vozzi}
\email{caterina.vozzi@ifn.cnr.it}  \affiliation{Istituto di Fotonica e Nanotecnologie - CNR, Milan, Italy}

\begin{abstract}
We investigated the giant resonance in Xenon by high-order harmonic generation spectroscopy driven by a two-color field. 
The addition of a non-perturbative second harmonic component parallel to the driving field breaks the symmetry between neighboring sub-cycles resulting in the appearance of spectral caustics at two distinct cut-off energies. By controlling the phase delay between the two color components it is possible to tailor the harmonic emission in order to amplify and isolate the spectral feature of interest.
In this paper we demonstrate how this control scheme can be used to investigate the role of electron correlations that give birth to the giant resonance in Xenon.
The collective excitations of the giant dipole resonance in Xenon combined with the spectral manipulation associated with the two color driving field allow to see features that are normally not accessible and to obtain a quantitative good agreement between the experimental results and the theoretical predictions.
\end{abstract}

\maketitle

High order harmonic generation (HHG) has proved to be a valuable spectroscopic tool for probing the electron structure \cite{Itatani_2004, Vozzi_2011_NP} and dynamics \cite{Li_2008, Smirnova_2009, Shafir_2012} of atoms and molecules. In this process an electron wave packet can be detached from the atom by tunnel ionization at each half-cycle of the driving laser field. This electron wave packet is then accelerated by the external field and can recollide with the parent ion, releasing its excess energy through the emission of an XUV burst \cite{Schafer_1993, Corkum_1993}. Each burst has the duration of a few hundreds of attoseconds, which makes HHG the cornerstone for the generation of the shortest events ever created.
The photon energy $\hbar\omega$ is linked to the time of birth of the electron in the continuum $t_i$ and to its recombination time $t_r$, defining a quantum trajectory \cite{Lewenstein_1994}. Since these processes occur on time scale shorter than the laser electric field cycle, and owing to the mapping between photon energies and electron trajectories, attosecond time resolution can be inferred through the analysis of HHG spectra.

One of the main challenges in Attosecond Science is the ability to manipulate the optical properties of the HHG spectrum, such as the spectral shape, polarization and phase. Such manipulations are the basic building blocks in a broad range of applications where the attosecond pulse served as a pump or a probe of a fast evolving process.
In this paper we show how an accurate manipulation of the HHG spectrum amplifies and isolates the spectral features associated with the attosecond dynamics under study.  When the HHG is generated by a two color field, an accurate control over its spectral shape can be achieved. Such control leads to the appearance of a spectral caustic \cite{Raz_2012}, allowing a tunable enhancement of a narrow spectral band. We engineered the HHG spectrum and tuned the caustic such that it overlaps with the spectral feature associated with the giant resonance in Xe, demonstrating for the first time how the application of such control serves as a valuable tool in HHG spectroscopy.

\begin{figure*}[h!]
\includegraphics[scale=0.4]{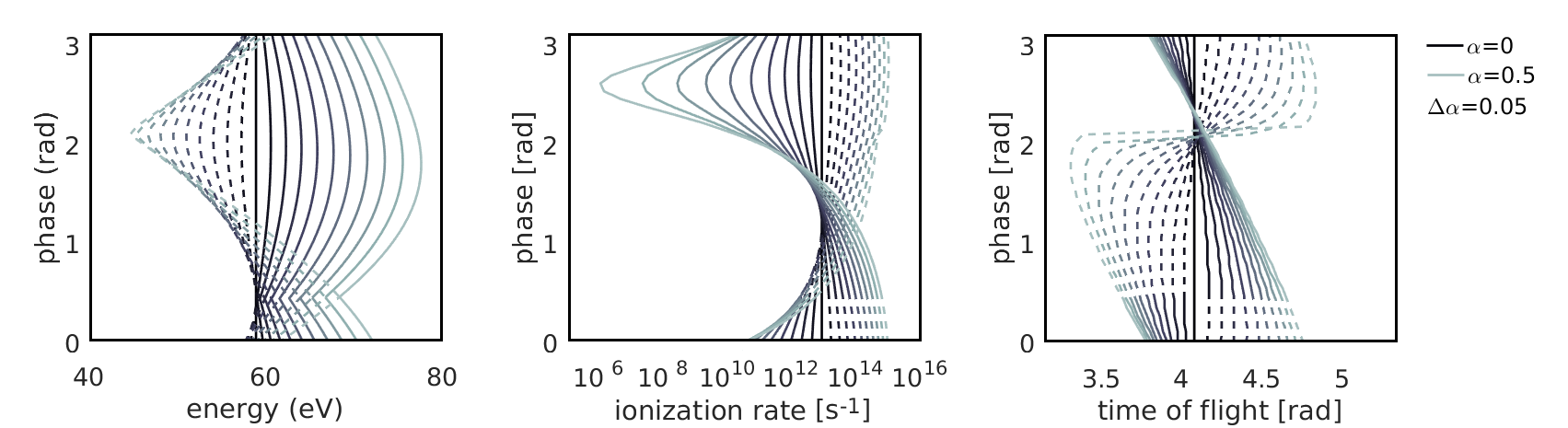}
\caption{(color online) Photon energies (a), ionization rates (b) and excursion times (c) associated to the cutoff trajectories as a function
of the phase difference between the two fields for different values of $\alpha$ ranging from 0 to 0.5 with step $\Delta\alpha=0.05$.
The dashed line represents the lower branch and the solid line represents the upper branch.
The two cutoff energies are calculated with the semiclassical three-step model; 
the ionization rates of the two cutoff trajectories are calculated with the ADK model \cite{Ammosov_1986};
the excursion time of the two cutoff trajectories is normalized to the period of the fundamental field.
The intensity of the fundamental field is equal to $I=7\times10^{13}\,{\mathrm{W}}/{\mathrm{cm}^2}$.}
\label{fig:1}
\end{figure*}

In a single color driving field, two trajectories contribute to the HHG signal for each photon energy, namely the \textit{short} and \textit{long} trajectories \cite{Lewenstein_1994}. The spectrum extends up to the cutoff energy $\hbar\omega_{\mathrm{cutoff}}= 1.32 I_p + 3.17 U_p$, where $I_p$ is the ionization potential of the target and $U_p$ is the ponderomotive energy of the free electron in the field. Approaching the cutoff, these two trajectories merge and only one trajectory is physically allowed \cite{Lewenstein_1994, Shan_2001}. 
When HHG is driven by the combination of the fundamental field with its second harmonic in parallel polarization, the symmetry between neighboring sub-cycles is broken and each trajectory is split in two distinct ones that recollide with the ion from opposite sides. This eventually leads to the presence of two distinct cut-off energies \cite{Figueira_de_Morisson_Faria_1999, Figueira_de_Morisson_Faria_2000, Frolov_2010}. Fig. \ref{fig:1}(a) shows the two semiclassical cutoffs as a function of the phase delay $\varphi$ between the two colors. The cutoff energy is shown for different values of the ratio between the two field components, $\alpha$, ranging from $0$ to $0.5$ with step $\Delta\alpha=0.05$. We can identify a \textit{lower branch} (lower cutoff, dashed line) and an \textit{upper branch} (upper cutoff, solid line). Approaching the cutoff the spectral density diverges to infinity \cite{Raz_2012}, leading to the appearance of a spectral caustic and allowing an enhancement of a narrow bandwidth in this spectral region. When more than two trajectories coalesce in the same spectral region this produces a dramatic enhancement as predicted by catastrophe theory. By changing $\varphi$ and $\alpha$, the appearance of the caustic can then be tuned to overlap with the spectral region of interest. However, in the experiment the visibility of the caustic also depends on the ionization probability and electron wave packet spreading of the associated cutoff trajectory. In Fig. \ref{fig:1}(b) and  \ref{fig:1}(c) the ionization probabilities and the excursion times of the recolliding electron for the two cutoff trajectories are reported. It is possible to observe that the lower branch has, for most of the values of $\varphi$, a higher ionization probability than the upper branch. In the remaining regions, namely for $\varphi$ approximately between $0.4\,\mathrm{rad}$ and $1.2\,\mathrm{rad}$, the excursion time is higher, which leads to a higher spreading of the electronic wavefunction and a lower recombination probability \cite{Figueira_de_Morisson_Faria_1999}. For these reasons, the upper branch has never been clearly observed in the experiments. However if the caustic overlaps a shape resonance of the target sample, the spectral focusing associated with the caustic combined with the enhanced cross section could efficiently counteract the reduction of the intensity making the full spectral features available for being probed.

As a benchmark case, we applied two-color HHG for probing the giant resonance in xenon, a very broad enhancement in the harmonic spectrum around 100 eV that has been predicted by Frolov \emph{et al.}  \cite{frolov_PRL_2009} and experimentally observed by Shiner \emph{et al.} \cite{Shiner_2011}. This enhancement was attributed to the multi-electron inelastic scattering that takes place during the recombination of the electron with the parent ion. The electron ionized from the $5p$ valence shell can recombine to the ground state by two possible channels. Either it can directly recombine with the $5p$ hole or it can exchange energy by inelastic scattering with one of the underlying $4d$ electrons, which is promoted to the $5p$ valence shell, leaving a hole in the $4d$ state. Eventually, the electron recombines with the $4d$ hole. This recombination channel can be accessed if the kinetic energy of the recolliding electron exceeds the energy difference between the two orbitals involved. This interpretation has been confirmed by time-dependent configuration-interaction singles (TDCIS) calculations taking into account the many-body interaction between different inner-shell orbitals~\cite{GrSa-PRA-2010,Pa-EPJST-2013}. TDCIS is a many-body theory that can capture multi-orbital effects in the attosecond~\cite{PaSa-PRL-2011,PaSy-PRA-2012} and strong-field regimes~\cite{PaGr-PRA-2012,PaSa-JPB-2014,PaSy-PRA-2012}, and has been successfully applied to capture the behavior of the collective dipole resonance in xenon in the presence of an intense XUV~\cite{PaWa-PRA-2015} and IR pulses~\cite{PaSa-PRL-2013}.

The quantitative comparison between experimental results and single atom predictions is usually very difficult.
One of the main problems arises from the fact that the intensity of the HHG spectrum can be shaped by phase matching effects \cite{Salieres_1995, Balcou_1997, Jin_2011}.
HHG is in fact the coherent build up of the single atom emissions of a macroscopic sample excited by a laser pulse.\cite{Vozzi_2011_NJP}.
For low pressure target media, phase-matching is essentially determined by the balancing between two terms:
the geometrical phase shift (\textit{Gouy phase}) and the 
dipole phase, which depends linearly on the driving intensity $I$, through a constant that takes different values according
to the trajectory considered  \cite{Balcou_1999,Lewenstein_1995}.
This leads to a different conversion efficiency for the two families of trajectories in the plateau.
Two color HHG is particularly favorable since at the caustic two trajectories coalesce in a single one, making it possible to analyze the resonance ruling out the interference of short and long trajectories that takes place in the plateau. Spectral focusing enhances the contribution of the cutoff trajectory with respect to the others, allowing to easily detect and probe a specific spectral region
that can be tuned around a wide spectral range.

The experimental setup is described in detail in \cite{Soifer_2014}.
The HHG driving pulse is a carrier envelope phase stabilized pulse at $\lambda_1=1550\,\mathrm{nm}$ with
a time duration of 25 fs \cite{Vozzi_2007}.
The second field component is the second harmonic of the fundamental pulse,
generated inside a $\beta$-Barium Borate crystal.
We used a calcite plate to correct the group delay between the two pulses and a birefringent retardation plate
for rotating the polarization of the fundamental field, thus switching to parallel polarization configuration.
A pair of fused silica wedges was used to fine tune the phase delay between the two laser fields.
The harmonics were generated by focusing the two color driving field into a pulsed gas jet.
In order to reduce the phase-mismatch of the harmonics in the cutoff region of the spectrum, where the Xenon giant resonance is located, the backing pressure of the gas behind the valve was kept below 1 bar.
The harmonic field was acquired by a XUV flat-field spectrometer \cite{Poletto_2001}
followed by a micro channel plate and a phosphor screen coupled to a CCD camera.

\begin{figure*}[htb]
\centering
\includegraphics[scale=0.5]{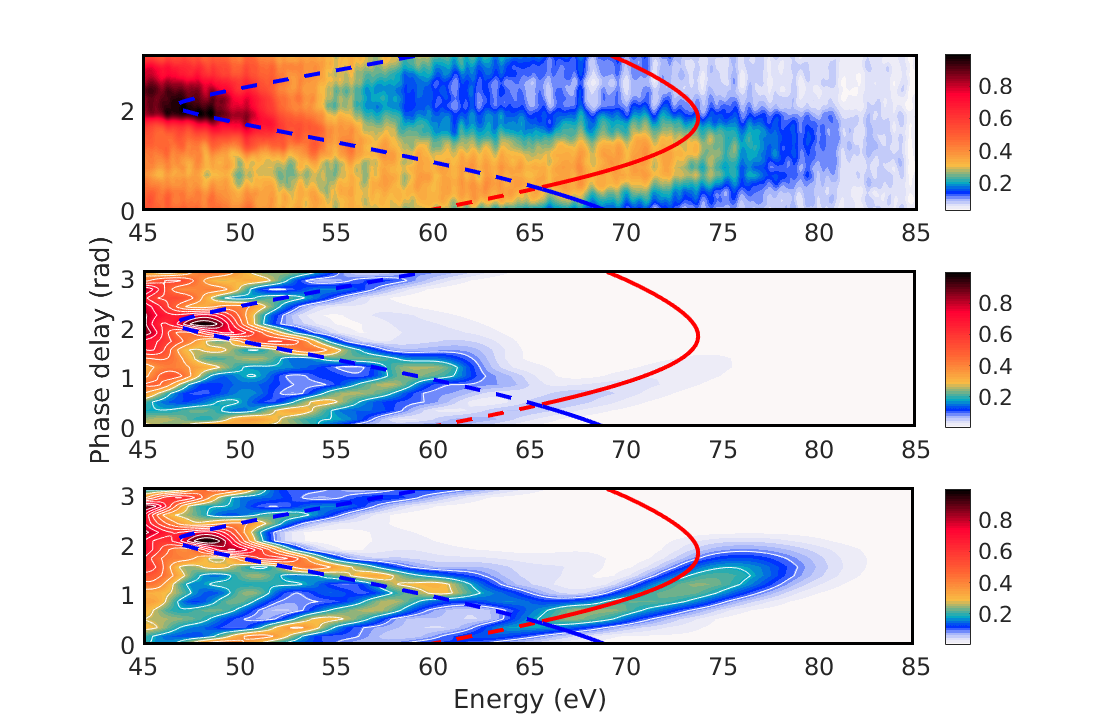}
\caption{(color online) HHG spectra as a function of the phase delay between the two colors - linear scale colormap. (a) Experiment
(b) TDCIS simulations with only the $5p_z$ orbital being active. 
(c) TDCIS simulations with all $5p$, $4d$ and $5s$ orbitals being active.
The solid and the dashed curves represent the two cutoffs for the upper and lower branches, respectively, associated with the semiclassical calculations.}
\label{fig:2}
\end{figure*}

Figure \ref{fig:2}(a) shows a sequence of harmonic spectra acquired in Xenon as a function of the two color phase delay $\varphi$.
Both caustics corresponding to the lower and the upper branches are clearly visible.
The strong enhancement of the lower caustic at $\varphi \approx\,$2.1 rad around 45 eV is due to the coalescence of 4 trajectories.
The unique behavior of the two caustics as a function of $\varphi$ and $\alpha$
allows the estimation of $\alpha$ and of the driving peak intensity of the fundamental color $I$ in the interaction region with a high-level of accuracy 
from the comparison with the semiclassical calculations in Fig. \ref{fig:1}.
The best fit corresponds to $I=7\times10^{13}\,{\mathrm{W}}/{\mathrm{cm}^2}$ and $\alpha= 0.404$.
The calculated cutoff according to the semiclassical model are reported on top of the color map.

Figure \ref{fig:2}(b-c) shows the calculated HHG spectrum based on TDCIS method with (b) only the $5p_z$ orbital active and (c) with all orbitals in the $4d$, $5s$, and $5p$ shells active.
HHG is calculated using the time-dependent dipole moment $\langle z \rangle (t)$. Restricting the manifold of active occupied orbitals to $5p_z$ mimics the single-active electron picture where the returning electron can only recombine with the originally occupied $5p_z$ orbital. 
Figure \ref{fig:2}(b) shows the upper branch is basically not visible in the HHG spectrum in the single-active electron (SAE) picture due to the spreading of the electron wavepacket and especially due to the reduced ionization probability of the corresponding electron trajectories. 
The experimental results in Fig.~\ref{fig:2}(a) show that major parts of the upper branch are visible, which is inconsistent with the SAE picture.
Including the $4d$, $5s$, and $5p$ shells and their interaction in the calculations leads to a very good agreement between theory and experiment [see Fig.~\ref{fig:2}(a,c)].
This upper branch has been hardly observed in other atoms where collective resonance effects are not present.
For example in Neon, Ishii \emph{et al.} \cite{Ishii_2008} observed a similar separation of the two cutoffs
performing two color driven HHG experiments with the fundamental wavelength of $800\,\mathrm{nm}$. 
However the conversion efficiency of the upper branch was very low in that case, due to the suppression already discussed in the single active electron picture.
Our results demonstrate that the enhanced recombination amplitude, due to the coupling between the $4d$ and $5p$ shells, counter acts the reduction in the tunnel ionization probability, and enables the measurement of spectral features that are normally inaccessible in HHG.

\begin{figure*}[htb]
\centering
\includegraphics[scale=0.5]{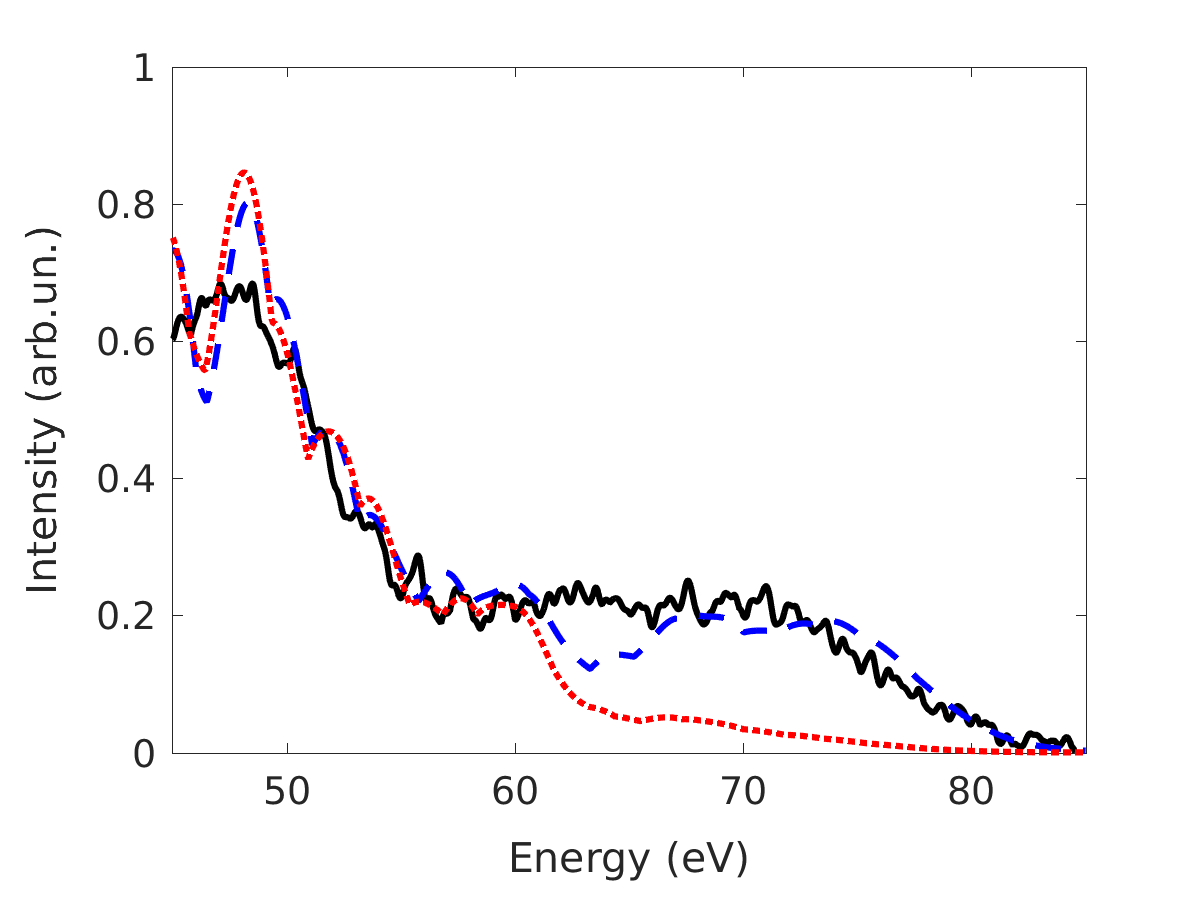}
\caption{(color online) Maximum value of HHG intensity with respect to the phase delay as a function of the photon energy. (solid line) Experiment.
(dotted line) TDCIS simulations considering only $5p$ orbitals.
(dashed line) TDCIS simulations including all $5p$, $4d$, and $5s$ orbitals.}
\label{fig:3}
\end{figure*}

In our measurements it is possible to clearly distinguish the two caustics of the upper and lower branch
and tune their spectral position with a high level of accuracy.
In these conditions is also possible to quantitatively compare the experimental results with the TDCIS calculations.
In Fig. \ref{fig:3} the maximum HHG intensity with respect to the phase delay $\varphi$ from the experimental data is shown as a function of the photon energy as solid line.
The predicted HHG intensities from TDCIS calculations with (dashed line) and without (dotted line) the contribution of the $4d$ and $5s$ orbitals are also shown.
The calculations are in very good agreement with the experimental data both from a 
qualitative and quantitative point of view only if the multi-electron effects are taken into account.
The aforementioned results show that the analysis of the caustics in non-perturbative two-color HHG can be considered a reliable probe for quantitatively comparing the experimental results with 
single-atom theories. Indeed in correspondence of the caustic the spectral focusing allows the detection of a specific trajectory that can be tuned around a wide spectral range.

In conclusion, we studied high-order harmonics in Xenon driven by a non-perturbative two-color field. With this technique,
we have shown how one can reveal and enhance a number of features in the HHG spectrum of Xe that are often hidden in other types of HHG spectroscopy.
The relatively strong second harmonic field breaks the symmetry between neighboring half-cycles leading to two cutoffs that are energetically well separated and enhanced by spectral focusing.
The upper branch, which is normally not visible in the experiments due to the strongly reduced tunneling and recombination rates, is clearly visible in xenon because of the collective excitation involving the $4d$,  $5s$, and $5p$ shells. The quantitative agreement of the experimental and theoretical results confirms that the upper branch is only visible due to the collective dipole excitation in xenon.
The analysis of the spectrum at the caustic offers a reliable probe for comparing the experiment with single atom predictions. We expect that this results can be extended beyond investigations of the structure of the giant resonance in Xenon  \cite{ChPa-PRA-2015}, and can pave the way toward more general HHG measurements of electron correlations and structural features such as shape resonances in many other atomic and molecular systems.

\begin{acknowledgements}
The research leading to these results has received funding
from the ERC Starting Research Grant UDYNI (grant agreement n 307964, EC
Seventh Framework Programme) and from the Italian Ministry of Research and Education (ELI project—ESFRI
Roadmap). S.P. is funded by the Alexander von Humboldt Foundation and by the
NSF through a grant to ITAMP.
N. D. acknowledges the Minerva Foundation, the Israeli Science Foundation,
the Israeli Centers of Research Excellence program, the Crown Photonics Center and the European Research Council Starting Research Grant MIDAS.
\end{acknowledgements}

\bibliography{Xe-2C.bib}

\end{document}